\documentclass[aps,prl,showpacs,twocolumn]{revtex4}
\usepackage[pdftex]{graphicx}
\usepackage{amssymb}
\usepackage{amsmath}
\usepackage{bm}

\DeclareGraphicsExtensions{.pdf}

\begin{document}

\title[]{Half-Quantum Vortices in an Antiferromagnetic Spinor Bose-Einstein Condensate}

\author{Sang Won Seo}
\author{Seji Kang}
\author{Woo Jin Kwon}
\author{Yong-il Shin}\email{yishin@snu.ac.kr}

\affiliation{Department of Physics and Astronomy, and Institute of Applied Physics, Seoul National University, Seoul 151-747, Korea}

\begin{abstract}
We report on the observation of half-quantum vortices (HQVs) in the easy-plane polar phase of an antiferromagnetic spinor Bose-Einstein condensate. Using {\it in situ} magnetization-sensitive imaging, we observe that pairs of HQVs with opposite core magnetization are generated when singly charged quantum vortices are injected into the condensate. The dynamics of HQV pair formation is characterized by measuring the temporal evolutions of the pair separation distance and the core magnetization, which reveals the short-range nature of the repulsive interactions between the HQVs. We find that spin fluctuations arising from thermal population of axial magnon excitations do not significantly affect the HQV pair formation dynamics. Our results demonstrate the instability of a singly charged vortex in the antiferromagnetic spinor condensate.
\end{abstract}


\maketitle

In a scalar superfluid, the supercurrent circulation around quantum vortices is quantized in units of $h/m$ due to $U(1)$ gauge symmetry~\cite{RJD}, where $h$ is the Planck constant and $m$ is particle mass. However, when a superfluid possesses an internal spin degree of freedom, there is an intriguing possibility for the superfluid to host quantum vortices of a fractional circulation of $h/m$. The superfluid phase interwinds with the spin orientation and a new relation is imposed on the supercurrent circulation in connection with spin texture~\cite{Volovik}.  Fractional quantum vortices are of particular interest in two-dimensional (2D) superfluidity. In the absence of long-range order in two dimensions~\cite{Mermin}, the superfluid phase transition in 2D is associated with vortex-antivortex pairing as described by the Berezinskii-Kosterlitz-Thouless (BKT) theory~\cite{Berezinskii,KT}. Hence fractional quantum vortices, introduced as new point defects, represent an interesting opportunity to explore for exotic superfluid phases, possibly beyond the BKT physics. 

Quantum vortices having $h/2m$ circulation, so-called half-quantum vortices (HQVs) have been experimentally observed in spinor superfluid systems such as exciton-polariton condensates~\cite{Lagoudakis,Manni,PNAS} and triplet superconductors~\cite{Jang}. In previous cold atom experiments, HQV states were created with an optical method in two-component Bose-Einstein condensates (BECs)~\cite{Cornell}, where the two components are not symmetric in terms of interactions. Recently, a  spin-1 BEC with antiferromagnetic interactions has been considered with great interest because HQVs are topologically allowed in the polar phase of the system~\cite{Zhou,Kawaguchi_review,Stamper-kurn_review}. Theoretical studies predicted an anomalous superfluid density jump at the phase transition in two dimensions~\cite{Chandra,Mukerjee,Pietila} as well as a new superfluid state that has completely broken spin ordering~\cite{Podolskii,Phase_Diagram}.

In this Letter, we report on the observation of HQVs in the easy-plane polar phase of an antiferromagnetic spinor BEC of $^{23}$Na atoms. Using magnetization-sensitive imaging, we observe that pairs of HQVs with opposite core magnetization are generated when singly charged quantum vortices are injected into the condensate. The temporal evolutions of the pair separation distance and the core magnetization reveal the short-range repulsive interactions between the HQVs. Dissociation dynamics of a singly charged vortex was previously observed in exciton-polariton condensates~\cite{Manni} but its dynamics was driven by spatially inhomogeneous spin-dependent potentials in the system. We emphasize that our system is spin symmetric and defect free. Thus, the observation of HQV pair formation clearly demonstrates the intrinsic instability of a singly charged quantum vortex in the spinor condensate. 

The spin-dependent part of the mean-field energy functional for a spin-1 spinor condensate is given as
\begin{equation}\label{eq1}
E_s=\frac{c_2 n}{2}\langle\mathbf{F}\rangle^2+p \langle F_z \rangle +q\langle F_z^2 \rangle,
\end{equation}
where $c_2$ is the spin-dependent interaction coefficient, $n$ is the atomic number density, $\mathbf{F}=(F_x,F_y,F_z)$ is the single-particle spin operator, and $p$ and $q$ are the linear and quadratic Zeeman fields, respectively~\cite{Kawaguchi_review,Stamper-kurn_review}. Here the external magnetic field defines the $z$ direction. For antiferromagnetic interactions ($c_2>0$) and zero total magnetization ($p=0$), the ground state of the system is a polar state with $\langle\mathbf{F}\rangle=0$~\cite{Ho,Machida}. This is a $|m_F=0\rangle$ state along a certain quantization axis which we denote by a unit vector $\vec{d}$. Depending on the sign of $q$, the condensate shows two distinctive phases: at $q>0$, the easy-axis polar phase with fixed $\vec{d}\parallel\hat{z}$, giving $\langle F_z^2\rangle=0$,  and at $q<0$, the easy-plane polar phase with $\vec{d}\perp\hat{z}$ and $\langle F_z^2\rangle=1$. In the following, we refer to these two phases as polar (P) and antiferromagnetic (AF) phases, respectively.

The order parameter of the AF phase can be expressed with a three-component spinor as
\begin{align}
\Psi_{\mathrm{AF}}=\left(
\begin{matrix}
  \psi_{+1}\\
  \psi_{0}\\
  \psi_{-1}
 \end{matrix}\right)=\sqrt{\frac{n}{2}}e^{i\theta}\left(
\begin{matrix}
  -e^{-i\phi}\\
  0\\
  e^{i\phi}
\end{matrix}\right),
\end{align}
where $\psi_l$ is the $m_z=l$ spin component along the $z$ direction ($l=0,\pm1$), $\theta$ is the superfluid phase, and $\phi$ is the spin orientation in the $xy$ plane, i.e. $\vec{d}=(\cos\phi,\sin\phi,0)$. The order parameter manifold is $\mathbb{M}_{\mathrm{AF}}=[U(1)\times S^1]/\mathbb{Z}_2$~\cite{Zhou,Mukerjee}, where $U(1)$ is the gauge symmetry, $S^1$ comes from the rotational symmetry of the spin, and $\mathbb{Z}_2$ arises from the invariance under the operation of $\theta\rightarrow \theta+\pi$ and $\phi\rightarrow \phi+\pi$. 

\begin{figure}
    \centering
    \includegraphics[width=7.8cm]{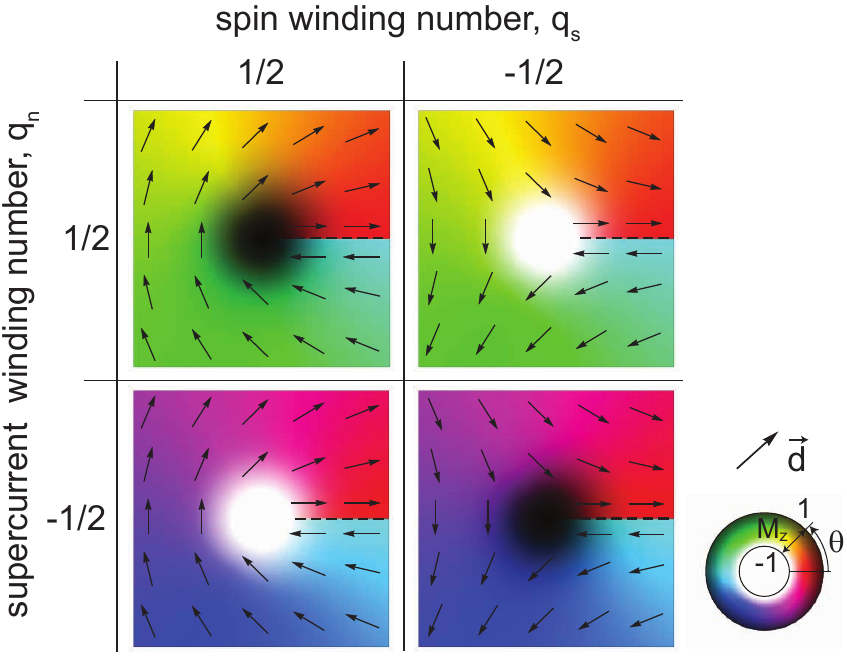}
    \caption{(color online). Schematic illustration of the half-quantum vortices (HQVs) in the antiferromagnetic (AF) spinor condensate. The superfluid phase $\theta$ and the spin orientation $\vec{d}$ rotate by $\pi$ around vortex cores having nonzero magnetization $M_z$. The order parameter of the condensate is invariant under the operation of $\theta\rightarrow \theta+\pi$ and $\vec{d}\rightarrow -\vec{d}$ and it is continuous over the disclinations indicated by dashed lines.}
    \label{fig1}
\end{figure}

When the windings of $\theta$ and $\phi$ around a quantum vortex are $q_n$ and $q_s$ in units of $2\pi$, respectively, the single-valuedness of the order parameter requires $q_n\pm q_s$ to be integer. Therefore, quantum vortices with a half-integer supercurrent winding number $q_n$ can exist with the aid of spin winding. The spatial structures of the four fundamental HQVs with $|q_n|=|q_s|=\frac{1}{2}$ are described in Fig.~1. When $q_n+q_s=0$ ($q_n-q_s=0$), the $m_z=-1$ ($m_z=1$) component has no vorticity and fills up the HQV core. Although a ferromagnetic core is costly for the AF spin interactions, the core filling is energetically favored as it would reduce the kinetic energy by suppressing the density of the circulating spin component in the core region.

A singly charged vortex with $(q_n,q_s)=(\pm 1,0)$ can be regarded as a sum of two HQVs : $(q_n,q_s)=(\pm\frac{1}{2}, \frac{1}{2})$ and $(\pm\frac{1}{2},-\frac{1}{2})$. It was predicted in mean-field calculations that in the AF phase a singly charged vortex state is energetically unstable to decay into two HQVs~\cite{HQVcoresize,Eto,Lovegrove}. Note that the two HQVs have opposite core magnetization. This means that if the disintegration of a singly charged vortex occurs in an AF spinor condensate, it would be identified with formation of a pair of ferromagnetic defects having opposite magnetization. 

\begin{figure}
\includegraphics[width=8.0cm]{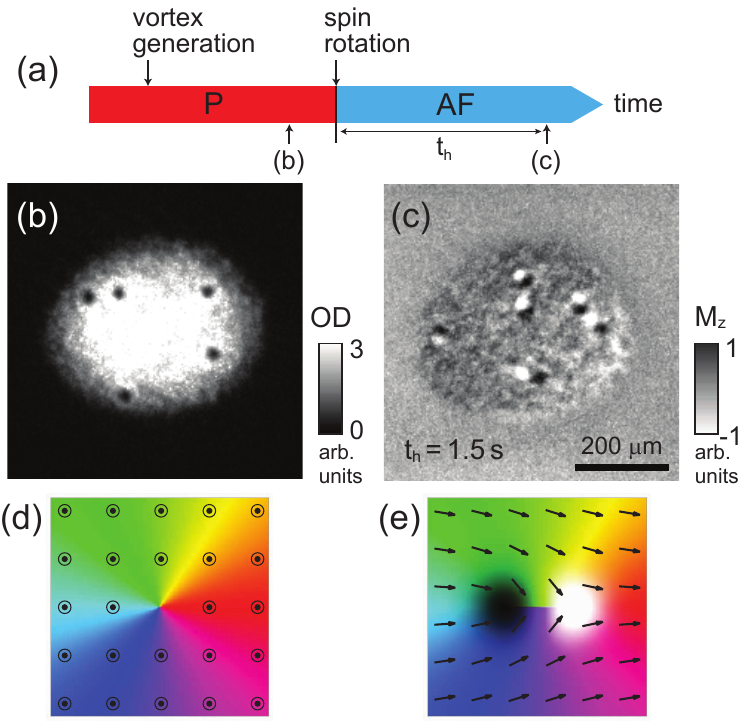}
\caption{(color online). Dissociation of a singly charged vortex into two HQVs. (a) Singly charged vortices are generated in a condensate in the polar (P) phase and then the condensate is transmuted to the AF phase (see the text). (b) Optical density (OD) image of a condensate containing singly charged vortices. (c) Magnetization-sensitive phase-contrast image of the AF condensate at $t_h=1.5$~s. The images were taken after 24-ms expansion by releasing the trapping potential. Schematic descriptions of (d) a singly charged vortex state with $(q_n,q_s)=(1,0)$ and (e) a state having a pair of HQVs with $(q_n,q_s)=(\frac{1}{2},\frac{1}{2})$ and $(\frac{1}{2},-\frac{1}{2})$.
}
\label{fig2}
\end{figure}

Our experiment starts by generating a BEC of $^{23}$Na atoms in the $|F=1,m_F=0\rangle$ hyperfine spin state in an optical dipole trap~\cite{Skyrmion_prl}. The trapping frequencies are $(\omega_x, \omega_y,\omega_z)/2\pi =(4.2, 5.3, 480)$~Hz and a typical condensate containing about $3.5\times 10^6$ atoms has the Thomas-Fermi radii $(R_x,R_y,R_z)\approx(185, 150, 1.6)~\mu$m.  For peak atom density, the spin healing length is $\xi_s=\hbar/\sqrt{2 m c_{2} n}\approx4.5~\mu$m~\cite{c2}, which is larger than the sample thickness $R_z$ and thus, the spin dynamics in the condensate is effectively 2D. The external magnetic field $B_z=30$~mG, giving $q/h=0.24$~Hz, and the residual field gradient is compensated to be less than $40 ~\mu$G/cm. 

We inject quantum vortices into the condensate by stirring the center region of the condensate with a repulsive laser beam for 10~ms, as described in Ref.~\cite{Kwon}. Because the condensate is in the $P$ phase with $U(1)$ symmetry and the multiply charged vortices are unstable to decay~\cite{MV}, it is ensured that the generated vortices are singly charged. The average vortex number was about six [Fig.~2(b)].

The condensate is transmuted into the AF phase by tuning the quadratic Zeeman field $q$ with the microwave dressing technique~\cite{Gerbier,Quenching,Quenching2,Zhao,Jiang}. First, we apply a radio-frequency pulse of 65~$\mu$s to rotate the spin from $+\hat{z}$ to $+\hat{x}$, forming a superposition state of the $m_z=\pm1$ components, and then we immediately turn on a microwave field with frequency detuned by $-300$~kHz with respect to the $|F=1,m_F=0\rangle \rightarrow |F=2,m_F=0\rangle$ transition, resulting in $q/h= -10$~Hz. We confirmed that the $m_z=0$ component is absent in the microwave dressing using Stern-Gerlach spin separation measurements.

The spatial distribution of the condensate magnetization is measured with a spin-dependent phase-contrast imaging method~\cite{Higbie,Carusotto,shin}. The probe light is circularly polarized and the frequency is detuned by $-20$~MHz from the $3S_{1/2}|F=1\rangle \rightarrow 3P_{3/2}|F^{\prime}=2\rangle$ transition~\cite{footnote1}. Because the phase shifts of the probe beam for the $m_z=\pm1$ components are opposite, the optical signal in the phase-contrast imaging is proportional to the density difference between the spin components, i.e., the magnetization $M_z=n\langle F_z\rangle$ of the sample.

\begin{figure}
\includegraphics[width=8.0cm]{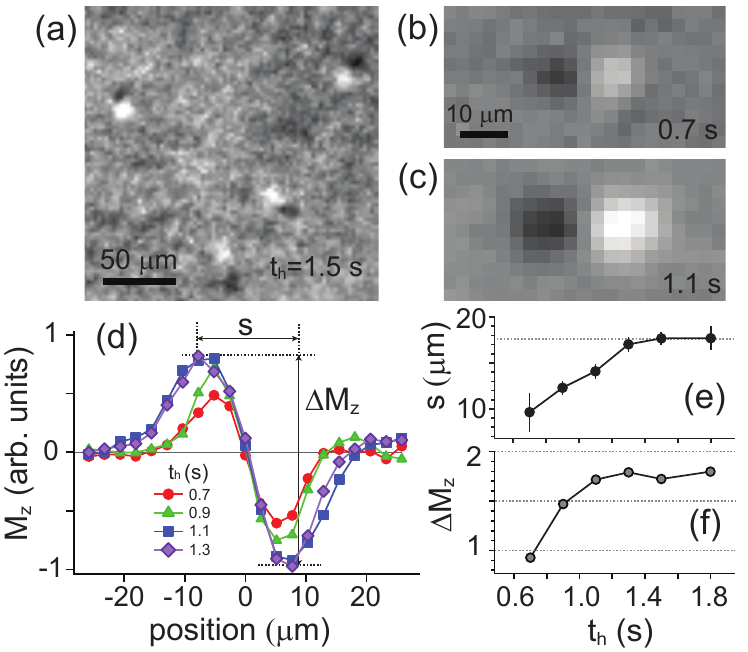}
\caption{(color online). Formation of a HQV pair with opposite core magnetization. (a) {\it In situ} magnetization distribution $M_z(x,y)$ of the condensate at $t_h=1.5$~s. Images of the HQV pairs at (b) $t_h=0.7$~s and (c) 1.1~s, obtained by averaging over ten images cropped from several magnetization images like (a). (d) Magnetization profiles of the HQV pairs along the separation direction at various hold times $t_h$, from averaged images like (b) and (c). Evolutions of (e) the pair separation distance $s$ and (f) the core magnetization difference $\Delta M_z$.
}
\label{fig3}
\end{figure}

When the condensate containing vortices is prepared in the AF phase, we observe that the vortex core visibility decreases in the optical density images like Fig.~2(b) when pairs of point defects with opposite magnetization emerge in the condensate~[Fig.~2(c)]~\cite{Supple}. The magnetic point defects are the HQV pairs that result from the dissociation of the singly charged vortices. We see that the separation directions of the HQV pairs are random, which clearly indicates that the dissociation dynamics is not driven by external perturbation such as the residual magnetic field gradient but by the intrinsic instability of a singly charged vortex state.

\begin{figure}
\includegraphics[width=7.6cm]{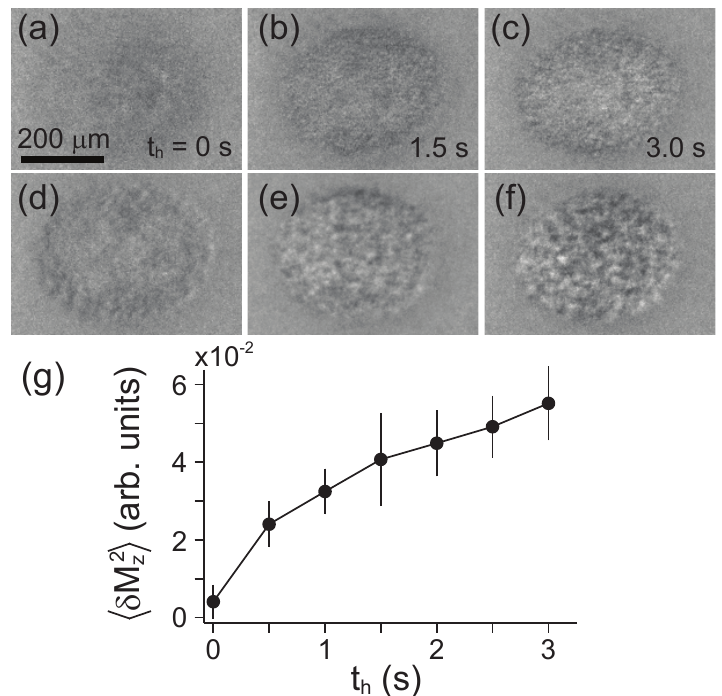}
\caption{Spin fluctuations in the AF spinor condensate. {\it In situ} magnetization images of the condensate, containing no vortices, at (a) $t_h=0$~s, (b) 1.5~s, and (c) 3~s. (d)-(f) Images taken after 24~ms time-of-flight, where spin fluctuations are enhanced due to self-interference during the expansion~\cite{Phase_Choi,Scaling}. (g) Variance of {\it in situ} magnetization $\langle \delta M_z^2 \rangle$ as a function of the hold time $t_h$. The background noise level is subtracted.}
\label{fig4}
\end{figure}

In {\it in situ} measurements of the magnetization distribution, HQV pairs become discernible after a hold time $t_h\sim 0.7$~s in the microwave dressing (Fig.~3). The separation distance of the pair and the magnitude of the core magnetization gradually increase, and the growth ceases after $t_h\sim1.3$~s when the separation distance reaches about $s_0= 17.6~\mu$m. This saturation behavior seems to be consistent with the short-range repulsive interactions between HQVs with opposite core magnetization~\cite{HQVcoresize,Eto}. Taking into account the imaging resolution of $\approx 4~\mu$m, we estimate the FWHM of a fully developed, magnetized HQV core to be $\approx 8.4$~$\mu$m that corresponds to $\sim 1.7\bar{\xi}_s$, where $\bar{\xi}_s$ is the average value of the spin healing lengths at the positions of the HQV pairs. This is in good quantitative agreement with mean-field predictions~\cite{HQVcoresize,Lovegrove}.

We observe that spatial fluctuations of magnetization develop in the condensate when it is transferred to the AF phase~(Fig.~\ref{fig4}). The AF condensate has two gapless excitation modes: phonons and axial magnons. Spin fluctuations arise mainly from thermal population of the magnon mode~\cite{Kawaguchi_review,Stamper-kurn_review,Symes}. In our experiment, when the condensate is initially prepared in the AF phase by the spin rotation, all the spins are aligned to the $+x$ direction and this spin texture corresponds to the zero temperature for magnon excitations. Thus, thermal relaxation of the spin temperature would subsequently occur and lead to spin fluctuations as observed. Because of the 2D character of spin dynamics, we may expect further enhancement of spin fluctuations in our system. Spin fluctuations show a steady increase after an initial, relatively rapid growth~[Fig.~4(g)], which we attribute to the heating effect of the microwave dressing~\cite{Jiang}.

\begin{figure}
\includegraphics[width=8.0cm]{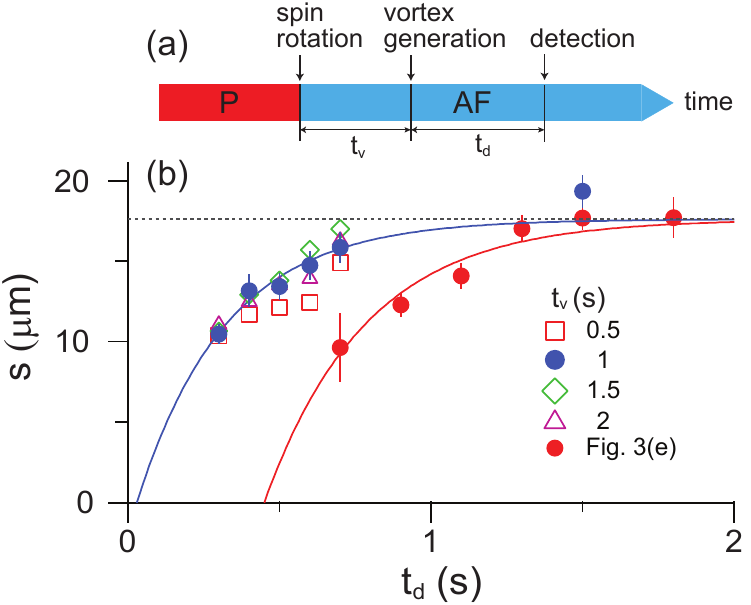}
\caption{(color online). Effect of thermal magnons on the pair dissociation. (a) Vortices are generated in the AF condensate after a dwell time $t_v$. Thermal spin fluctuations increase with $t_v$ [Fig.~4(g)]. (b) Pair separation distance $s$ as a function of the hold time $t_h$ after vortex generation. The data in Fig.~3(e) are displayed together and labeled as $t_v=0$~s. The solid lines are the lines of $s_0(1-e^{-(t_d-t_0)/\tau_s})$ with $s_0=17.6~\mu$m, fit to the data: $(t_0,\tau_s)=(0.03, 0.29)$~s for $t_v=1$~s (blue) and $(t_0,\tau_s)=(0.44, 0.35)$~s for $t_v=0$~s (red). 
}
\label{fig5}
\end{figure}

The dissociation of a singly charged vortex involves spin texture formation as well as magnetized core development, as depicted in Fig.~2(e). Therefore, magnons that are spin wave excitations might affect the pair dissociation dynamics. To investigate the possible effect of thermal magnons, we measure the temporal evolution of the pair separation distance at various spin temperatures of the condensate. To vary the spin temperature, we exploit the aforementioned thermal relaxation of magnons and let the condensate dwell in the microwave dressing for a time $t_v$ before generating vortices by stirring~[Fig.~5(a)].  Here we assume that HQVs cannot be directly generated by the stirring because the moving optical potential induces only density perturbations in the condensate. 

We find that the increasing rate of the pair separation distance is not significantly affected by thermal spin fluctuations. This implies that once a singly charged vortex is split, the subsequent dynamics of HQV pair formation is mainly driven by the repulsive interactions of the HQVs. The characteristic time scale of the pair formation is estimated by fitting a growth model $s_0 (1-e^{-(t_d-t_0)/\tau_s})$ to the experimental data, which results in $\tau_s\sim 0.3$~s, where $t_d$ is the hold time after vortex generation [Fig.~5(b)]. We note that the value of $\tau_s$ is roughly comparable to the time scale set by the spin-dependent interaction energy $c_2 n/h\sim 10$~Hz. The model fitting to the data for $t_v>0.5$~s and the data of the previous experiment [red circles in Fig.~5(b)] gives  $t_0\sim 0$~s and 0.5~s, respectively. This seems to suggest that thermal spin fluctuations at $t_v>0.5$~s are large enough to quickly split the vortex to a certain separation distance beyond which the repulsion between the HQVs becomes prominent.

In conclusion, we have observed HQVs with magnetized cores and confirmed the instability of a singly charged vortex in the AF spinor condensate. We will extend this work into a 2D regime~\cite{BKTcossover,Choi}, where a pair superfluid state without spin ordering is predicted to exist at finite temperatures~\cite{Podolskii,Phase_Diagram}. The spin ordering of the system might be probed spin-sensitive Bragg spectroscopy~\cite{Bragg,Veeravalli} or matter wave interference methods~\cite{Clade}.

We thank Fabrice Gerbier for advice on the microwave dressing and Suk Bum Chung for valuable discussion. This work was supported by the National Research Foundation of Korea (Grants No. 2011-0017527 and No. 2013-H1A8A1003984).


\begin{references}
\bibitem{footnote1} Because the probe light is not far from the resonance, the phase-contrast imaging involves non-negligible absorption effects and we empirically tune the probe light frequency to obtain a most flat signal for the condensate in the AF phase.
\end{references}
\end{document}